\documentclass[twocolumn,preprintnumbers,amsmath,amssymb]{revtex4}
\usepackage{graphicx}
\usepackage{amssymb}

\begin{document}

\title{Measurement-device-independent quantum key distribution with
classical Bob and no joint measurement}
\author{Guang Ping He}
\email{hegp@mail.sysu.edu.cn}
\affiliation{School of Physics, Sun Yat-sen University, Guangzhou 510275, China}

\begin{abstract}
Measurement-device-independent quantum key distribution (MDI-QKD) provides a
method for secret communication whose security does not rely on trusted
measurement devices. In all existing MDI-QKD protocols, the participant
Charlie has to perform the Bell state measurement or other joint
measurements. Here we propose an MDI-QKD protocol which requires individual
measurements only. Meanwhile, all operations of the receiver Bob are
classical, without the need for preparing and measuring quantum systems.
Thus the implementation of the protocol has a lower technical requirement on
Bob and Charlie.
\end{abstract}

\maketitle


\section{Introduction}

Quantum key distribution (QKD) is known to be a method for two parties Alice
and Bob to exchange classical secret information by transmitting quantum
states, usually encoded using photons \cite{qi365}. In principle, the
security of QKD against eavesdropping can be based solely on the validity of
the axioms of quantum mechanics. But in practice, such an unconditional
security is much harder to achieve, due to the imperfection of the actual
devices used in the implementation schemes \cite{qi1399}. Especially, the
theory-practice deviation of photon detectors could leave room for
side-channel attacks and blinding attacks \cite{qi863,qi1005,qi1979}.

Then came the measurement-device-independent (MDI)-QKD \cite{di22} as a
solution. It allows the detectors to be handled by a third party Charlie,
who cannot learn the secret information of Alice and Bob even if he himself
is the eavesdropper. That is, the security of MDI-QKD does not need to rely
on the assumption that the detectors are controlled by honest parties. Note that it does not necessarily mean that MDI-QKD has to be a three-party cryptography. Instead, when there are only Alice and Bob, and the latter owns the detectors, then \textquotedblleft Charlie\textquotedblright\ can be understood as the backdoor or spyware built into the detectors, which can communicate with or be controlled remotely by the eavesdropper. The existence of unconditionally secure MDI-QKD protocols mean that Bob can use the detectors from any manufacturer, even those from his enemy, while his secret information will remain secure.
On the other hand, comparing with full device-independent QKD (see Refs. \cite%
{qi481,qi996,qi997,qi998,di3,di21} and the references therein) which is
secure even when the eavesdropper can access to not only the detectors but also other devices, MDI-QKD is
much more feasible and the key rate can be very high. Therefore, it
immediately caught great interests \cite%
{di23,di30,di82,di93,di130,di141,di178}.

But the measurements in MDI-QKD are more complicated than those in previous
conventional ones. In some of the conventional QKD such as the
Bennett-Brassard84 (BB84) protocol \cite{qi365}, Bob can simply perform
individual measurements on every single photon from Alice one by one. In
MDI-QKD such as the protocol in Ref. \cite{di22}, however, both Alice and
Bob send photons to Charlie, and Charlie needs to perform the Bell state
measurement on both photons simultaneously. In a recent variation of MDI-QKD
called the twin-field (TF) MDI-QKD \cite{qi1517,qi1798,qi1956}, both Alice
and Bob send weak coherent pulses and Charlie measures the interference when
they combine on a beam splitter. Either way, Charlie's operations are joint
measurements. There was an MDI-QKD protocol claimed to be free from joint
measurements \cite{di479}, but its measurement for Charlie is actually a
joint measurement on two pulses sent to him at different times. While all
the above joint measurements are experimentally available with
state-of-the-art technology, they are undoubtedly less efficient and
convenient than individual measurements. Also, as pointed out in Ref. \cite%
{di479}, most MDI-QKD protocols have to deal with the difficulty of the
synchronization of the arrival times and phases of the photons or pulses from Alice and
Bob.

In this paper, we will propose an MDI-QKD basing on a completely different
route, so that strictly no joint measurement is needed. This makes it
possible to enjoy the advantage of MDI (i.e., the protocol remains secure
even when the measurement devices are controlled by the eavesdropper) using
conventional optical detectors for individual measurements, which are generally more efficient than those for joint measurements. Moreover, our protocol also
has an intriguing feature that the operations of Bob can all be considered
classical.

The structure of the rest of this paper is as follows. In the next section,
the general theoretical description of our protocol will be given, with its
security discussed in section III. Then in section IV, we study a possible
practical implementation of the protocol. The reason why Bob can be
classical is elaborated in section V. Finally we summarize the result and
compare its pros and cons with existing MDI-QKD protocols.

\section{The theoretical scheme}

Unlike previous MDI-QKD protocols where Alice and Bob both send photons or
coherent pulses for Charlie to measure, in our proposal only Alice sends
photons to Charlie, while Bob performs some secret operations on these
photons in the middle. To make the theoretical security analysis more
explicit, here we first give a general description of our MDI-QKD protocol
without going into the details on the implementation of the carriers of the
quantum states and how they are transmitted, and consider the ideal case where the transmissions and detections are free from any loss and error.

\textbf{Our protocol:}

(1) For $i=1$ to $n$:

\qquad (1.1) Alice sends Bob a batch of $m$ quantum registers $\psi _{1}^{(i)}$, $\psi _{2}^{(i)}$, ..., $\psi _{m}^{(i)}$ where $m$ is an even
number and $m\geq 8$ is recommended. Each register is a two-level system, whose state $\left\vert \psi
_{j}^{(i)}\right\rangle $\ ($j=1,...,m$) is randomly chosen from $%
\{\left\vert 0\right\rangle ,\left\vert 1\right\rangle ,\left\vert
+\right\rangle ,\left\vert -\right\rangle \}$. Here $\left\vert
0\right\rangle $ and $\left\vert 1\right\rangle $\ are two orthogonal states
of the two-level system, and $\left\vert \pm \right\rangle \equiv
(\left\vert 0\right\rangle \pm \left\vert 1\right\rangle )/\sqrt{2}$.

\qquad (1.2) Bob rearranges the order of the quantum registers. That is, he
chooses randomly a permutation operation $P_{i}$ and applies it on $\psi _{1}^{(i)}$, $\psi _{2}^{(i)}$, ..., $\psi _{m}^{(i)}$ to obtain $\left\vert \phi
_{1}^{(i)}\phi _{2}^{(i)}...\phi _{m}^{(i)}\right\rangle =P_{i}\left\vert
\psi _{1}^{(i)}\psi _{2}^{(i)}...\psi _{m}^{(i)}\right\rangle $. Bob keeps
his choice of $P_{i}$ secret, which will never be announced throughout the
entire protocol. Then he sends $\phi _{1}^{(i)}$, $\phi _{2}^{(i)}$, ..., $\phi
_{m}^{(i)}$ to Charlie, who is in charge of the measurement devices.

\qquad (1.3) Charlie is supposed to measure each of $\phi _{1}^{(i)}$, ..., $\phi
_{m/2}^{(i)}$ in the rectilinear basis $\{\left\vert 0\right\rangle
,\left\vert 1\right\rangle \}$, and measure each of $\phi
_{m/2+1}^{(i)}$, ..., $\phi _{m}^{(i)}$ in the diagonal basis $\{\left\vert
+\right\rangle ,\left\vert -\right\rangle \}$. Then he announces the
measurement results to Bob.

\qquad (1.4) Since Bob himself knows the permutation $P_{i}$, he deduces the
measurement result of each of $\psi _{1}^{(i)}$, $\psi _{2}^{(i)}$, ..., $\psi
_{m}^{(i)}$ from Charlie's announced information. Then they repeat steps
(1.1)-(1.4) for the next $i$, where Bob should choose a different $P_{i}$ in
step (1.2).

(2) Alice announces the bases (but not the exact state information) of all
the $n\times m$ registers $\psi _{j}^{(i)}$\ ($i=1,...,n$, $j=1,...,m$).

(3) The security check: Bob picks a portion of the registers and asks Alice
to announce their exact states. Then he checks whether Charlie's measurement
results matches Alice's announced states whenever Charlie has measured the corresponding
registers in the correct basis (i.e., he has measured $\left\vert
0\right\rangle $, $\left\vert 1\right\rangle $ ($\left\vert +\right\rangle $%
, $\left\vert -\right\rangle $) in the rectilinear basis (the diagonal
basis)), or the two states in the same basis occur with approximately equal
probabilities whenever Charlie has measured the registers in the wrong basis.

(4) If no suspicious result is found, among the rest registers which were
not picked for the security check, Bob keeps those which Charlie has
measured in the correct basis, and announces their indices $i$, $j$ to
Alice. Since both Alice and Bob know the exact state of this portion of
registers, they take $\left\vert 0\right\rangle $, $\left\vert
+\right\rangle $\ as the bit $0$\ and $\left\vert 1\right\rangle $, $%
\left\vert -\right\rangle $\ as the bit $1$, and thus obtain the raw secret
key.

\section{Security analysis}

We have to admit that at the present moment, we are unable to give a
rigorous mathematical security proof of the protocol which could be
sufficiently general to cover any cheating strategy that may potentially
exist, like the proofs in Refs. \cite{qi67,qi70} for the BB84 protocol. And
we wish that, like the case of the original MDI-QKD protocol \cite{di22},
related rigorous proofs can be eventually completed by successive studies
from various contributors \cite{di31,di233,di262,di292,di323,di340,di374}.
But for now, at least we can obtain the following conclusions in a heuristic
way.


\textbf{Theorem 1.} Without knowing Bob's permutation operations, Charlie's
announcing the measurement result dishonestly will be discovered with a
non-trivial probability.

Proof: In our protocol, Charlie is required to announce the measurement
result in step (1.3), before Alice announces the bases of the quantum states
in step (2). Therefore, he cannot delay his announcement until he learns the
basis information. For each $i$, there are totally $m!$\ possible choices
for the $m$-body permutation operation $P_{i}$. When Charlie does not know $P_{i}$,
for each specific $\phi _{j}^{(i)}$, form his point of view it could be any
of $\psi _{1}^{(i)}$, $\psi _{2}^{(i)}$, ..., $\psi _{m}^{(i)}$. Even if he
intercepted so that he owns or has owned all $\psi _{1}^{(i)}$, $\psi
_{2}^{(i)}$, ..., $\psi _{m}^{(i)}$, there is no measurement on $\psi
_{1}^{(i)}$, $\psi _{2}^{(i)}$, ..., $\psi _{m}^{(i)}$\ that can help him
know the correct basis for $\phi _{j}^{(i)}$. Consequently, no matter he
measures $\phi _{j}^{(i)}$ as required by the protocol or delays his
measurement while announcing a random measurement result, his announced
basis stands probability $1/2$ to be correct. In this case, if he announces
a result opposite to what is obtained in his measurement (or announces a
random result without actually performing the measurement) as the state of $%
\phi _{j}^{(i)}$, there is probability $\varepsilon _{ij}=1$ ($\varepsilon
_{ij}=1/2$) that it will conflict with the actual state prepared by Alice.
Once this $\phi _{j}^{(i)}$ is picked for the security check in step (3),
Bob will discover such a cheating. If Charlie totally did this for $k$
registers, then the probability for him to pass the security check will be
at the order of magnitude of $(1-\varepsilon /2)^{\lambda k}$, which drops
exponentially as $k$ increases. Here $\varepsilon \in \lbrack 1/2,1]$\ and $%
\lambda $ marks the portion of the registers that are picked for the
security check and Charlie's announced measurement bases happen to be
correct.

\bigskip

The above theorem guarantees that the performance of the measurement devices in
the protocol is checkable, so that Alice and Bob can arrive at the same raw
key correctly.

Meanwhile, the result below also holds.


\textbf{Theorem 2.} Without knowing Bob's permutation operations, the
eavesdropper (including Charlie) cannot obtain a non-trivial
amount of information on the secret key of Alice and Bob.

Proof: According to step (4), to learn a single bit of the raw key, the
eavesdropper must know whether the state of Alice's corresponding quantum
register $\psi _{j}^{(i)}$ belongs to $\{\left\vert 0\right\rangle
,\left\vert +\right\rangle \}$ or $\{\left\vert 1\right\rangle ,\left\vert
-\right\rangle \}$. Theorem 1 ensures that Bob can deduce the correct
measurement result of $\psi _{j}^{(i)}$. So if we treat Bob and all
measurement results as a whole party, then the situation between Alice and
Bob is actually the same as that of the BB84 protocol. As a result, it is
not hard to see that any intercept-resend attack on the quantum transmission
channel between Alice and Bob cannot pass the security check with a
non-trivial probability. That is, the eavesdropper (no matter he is Charlie
himself or someone else) cannot intercept and measure Alice's state before
it enters Bob's site. On the other hand, it is indeed possible to eavesdrop
the measurement result of $\phi _{j}^{(i)}$ (by either intercept the state
at Bob's output or get the information directly from Charlie). But as long
as Bob's permutation $P_{i}$ is kept secret, the relationship between $\psi
_{j}^{(i)}$\ and $\phi _{j}^{(i)}$ cannot be deduced. That is, eavesdropping
at Bob's output and/or Charlie's site is insufficient for deducing the
secret bits either.

\bigskip

Putting theorems 1 and 2 together, we can see that the hinge to the security
of the protocol is to keep Bob's permutations $P_{i}$ unknown to the
eavesdropper. According to the protocol, Bob chooses every $P_{i}$ by
himself, and applies it locally in his own site without announcing anything
about it throughout the protocol. Therefore, it seems easy to meet this
requirement in principle.

Nevertheless, in practice it is more complicated. The implementation details
of how the quantum states are transmitted may leave rooms for the
eavesdropping. To be precise, when the eavesdropper takes control on both
the input and output of Bob's site and is able to replace Alice's states
with something else he prepared himself, it is possible for him to learn
what operation is applied within Bob's site. Note that each $P_{i}$ is an $m$%
-qubit permutation operation. When describing Alice's $m$-qubit state as a $2^{m}$%
-dimensional\ vector, $P_{i}$ is corresponding to a $2^{m}\times 2^{m}$\
matrix. As there are $m!$\ possible choices for $P_{i}$, its matrix has $m!$%
\ independent elements. Denote $\left\vert \eta ^{(i)}\right\rangle $\ as
the state that the eavesdropper inputs to Bob's site. Then the output state
is $P_{i}\left\vert \eta ^{(i)}\right\rangle $. The eavesdropper's task is
to deduce $P_{i}$ by measuring $P_{i}\left\vert \eta ^{(i)}\right\rangle $.
To determine $P_{i}$ unambiguously, $P_{i}\left\vert \eta
^{(i)}\right\rangle $ should be orthogonal to $P_{i^{\prime }}\left\vert
\eta ^{(i^{\prime })}\right\rangle $ for any $i^{\prime }\neq i$, i.e.,%
\begin{equation}
\left\langle \eta ^{(i^{\prime })}\right\vert P_{i^{\prime }}^{\dag
}P_{i}\left\vert \eta ^{(i)}\right\rangle =0.
\end{equation}%
But as long as $m\geq 4$, there is%
\begin{equation}
m!>2^{m}.
\end{equation}%
Thus it is clear that using a $2^{m}$-dimensional\ system as $\left\vert
\eta ^{(i)}\right\rangle $\ (which has $2^{m}$ orthogonal states only) is
insufficient for determining $P_{i}$ unambiguously from all the $m!$\
possible choices. Instead, the eavesdropper has to use a higher-dimensional
quantum system. Consequently, there is a difference in the dimension of the
physical systems between the eavesdropper's $\left\vert \eta
^{(i)}\right\rangle $ and Alice's actual $\left\vert \psi _{1}^{(i)}\psi
_{2}^{(i)}...\psi _{m}^{(i)}\right\rangle $ to Bob's input. The question is
whether Bob can tell the difference in practice.

A safe bet is to use the \textquotedblleft teleportation
trick\textquotedblright : Bob does not allow the physical carriers of Alice's
states to enter his site directly. Instead, he uses quantum teleportation
\cite{qi179} to transfer the quantum information of Alice's states to his
own physical systems, whose dimension is completely under his control.
With this method, he can be sure that the eavesdropper cannot fake Alice's
physical systems with other systems and use them as a Trojan horse device to
learn the information on his operation $P_{i}$ (see Ref. [6] of Ref. \cite%
{di22} for details). So we can indeed achieve an unconditionally secure
implementation of our protocol in practice. However, the quantum
teleportation procedure requires Bob to perform additional measurements,
making the implementation more inconvenient. Moreover, handing these
measurement tasks to Charlie may cause extra security problems, while
letting Bob himself to perform the measurements will make the protocol no
longer an MDI one. Thus, to achieve both unconditional security and the MDI
property simultaneously, we must take extra care on how the protocol is
actually implemented in practice.

\section{A possible practical implementation}


\begin{figure}[b]
\includegraphics[scale=0.69]{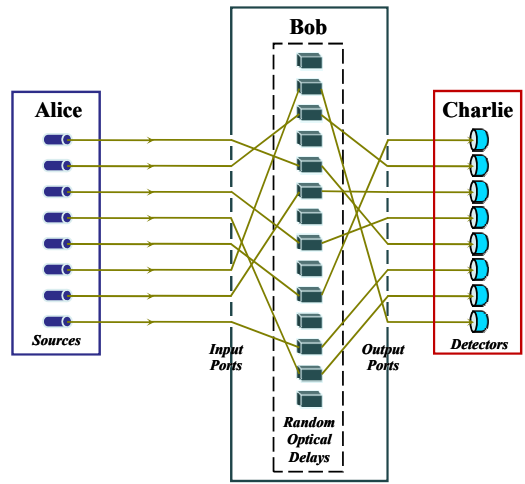}
\caption{Diagram of a practical implementation scheme of our MDI-QKD
protocol with $m=8$. All green lines stand for optical fibers. The
combination of the optical fibers inside Bob's site should be arranged
differently in each $i$th ($i=1,...,n$) round of the protocol.}
\label{fig:epsart}
\end{figure}


In Fig.1, we illustrate a possible implementation scheme of our protocol,
where Alice's single-photon sources, Bob's optical delays and Charlie's
detectors are all connected via optical fibers. The whole duration of the
quantum communication in the protocol is divided into $n$ time slots. In
each $i$th slot ($i=1,...,n$), Alice sends a photon from each of the $m$
sources, with the polarization direction prepared randomly as $0^{\circ }$, $%
90^{\circ }$, $45^{\circ }$ or $135^{\circ }$, which stand for the states $%
\left\vert 0\right\rangle ,\left\vert 1\right\rangle ,\left\vert
+\right\rangle $ or $\left\vert -\right\rangle $, respectively. She does not
need to send these $m$ photons at exactly the same moment. As long as their
sending times belong to the same time slot then it will be fine. All photons
then pass through Bob's site and finally reach Charlie's detectors. Bob
should choose a different combination of the optical fibers between his
input and output ports for each time slot, which serves as choosing a
different permutation operation $P_{i}$ on the $m$ photons. For easy
analysis and understanding on the security, we drew $m=8$ pairs of optical
sources and detectors in Fig.1. But in practice, by using the time division
multiplexing technique \cite{qi2018,qi2019}, only one source and one
detector will be sufficient.

Like many other practical QKD systems, Bob should also use single-mode
optical fibers, wavelength filters and single-photon filters \cite{qi2024}
right after his input ports, to ensure that photons with extra dimensions
or other characterizing information cannot enter his site \cite{qi1979}, and
the eavesdropper cannot input many photons to a single port simultaneously.
Otherwise, in every time slot the eavesdropper can simply prepare and send
different numbers of photons into different input ports of Bob. For example,
he sends $1$ photon to the first input port, and $2$ photons to the second
input port ... . Then by measuring the photon number at each of Bob's output
port, he can deduce Bob's permutation $P_{i}$. Meanwhile, he intercepts all
Alice's photons, applies $P_{i}$ on them and resends them to Charlie. As
long as all these operations can be completed fast enough, he can hack the
protocol without being revealed.

Nevertheless, though the use of the above filters can prevent the
eavesdropper from inputting many photons to one input port simultaneously,
he can still send these photons one by one. To avoid such a cheating, Bob
should shut down all input ports before and after each time slot. Meanwhile,
it is recommended to use low speed optical fibers in Bob's site (or at least use a short length of them after each input port) instead of
better ones, to serve as a time-based photon number filter. Suppose that his optical fibers can transmit $x$ bits per
second, while the duration of each time slot is $\tau $ seconds. Then Bob
should better choose optical fibers with speed%
\begin{equation}
x<<m/\tau .
\end{equation}%
In the most ideal case, when there is $x<2/\tau $, we can be sure that no
eavesdropper can input more than one photon to an input port during a time
slot. This may seem technically challenging, because it means that the
duration of each time slot should be as short as%
\begin{equation}
\tau <2/x,  \label{duration}
\end{equation}%
which implies that Bob needs very high speed switches to toggle the input
ports on and off. But if the time division multiplexing technique is used
for the implementation of our protocol, a single time slot will be shared by all
the $m$ ports. Then Eq. (\ref{duration}) will be replaced by%
\begin{equation}
\tau <2m/x.
\end{equation}%
We can see that it becomes possible to find suitable switches for the
implementation when $m$ is high.

For simplicity, the aforementioned filters and shutters are not shown in
Fig.1. Other than that, when comparing with the above theoretical
description of our protocol, the most distinct feature of Fig.1 is the
presence of the optical delays in Bob's site. They are adopted against
eavesdropping too. This is because, when these optical delays are not
present and Bob connects the input ports to the output ports in random order
using optical fibers directly, the eavesdropper (regardless he is Charlie or
not) can still learn Bob's $P_{i}$ using the following strategy. In each
time slot, though Bob's filters and shutters limit the eavesdropper to input
a single photon to each port only, each photon can be input at a slightly
different time. That is, the eavesdropper can send a photon into Bob's first
input port at time $t_{1}$, another photon into Bob's second input port at
time $t_{2}$, a third photon into Bob's third input port at time $t_{3}$,
... , and a photon into Bob's $m$th input port at time $t_{m}$. Here $%
t_{1}<t_{2}<...<t_{m}$\ and they are all within the same time slot, while
their difference is much smaller than the duration time $\tau $. The eavesdropper then
performs measurements at Bob's output ports to see when will a photon be
detected at each port. Through the time order of the photon detection, he
can deduce Bob's permutation $P_{i}$ even though he sends only one photon to
each port.%
To defeat this attack strategy, we need the optical delays shown in the
dashed box of Fig.1. The number of these delays should be larger than $m$,
and each of them is set to a different delay time. Then Bob's randomizing
the connection between these optical delays and the input ports is
equivalent to introducing a random delay time to the photon from each port.
In this case, if the eavesdropper applies the above attack by sending $m$
photons in sequence, these photons will leave Bob's output ports in a randomized
time order, so that the eavesdropper can no longer deduce which is the input
port that each photon was sent from, making the attack futile.

\section{Classicality of Bob}

The concept \textquotedblleft QKD with classical Bob\textquotedblright\ was
previously proposed by Boyer, Kenigsberg and Mor \cite{qi443}. In their
protocol, Bob's technical capability is limited to the following three
operations: (1) measuring the qubit from Alice in the computational basis $%
\{\left\vert 0\right\rangle ,\left\vert 1\right\rangle \}$, (2) preparing a
(fresh) qubit in the computational basis and sending it, and (3) reflecting
the qubit from Alice back undisturbed. The authors elaborated that such
operations can be considered classical, because unlike conventional QKD
(e.g., the BB84 protocol) that uses both the measurement bases $\{\left\vert
0\right\rangle ,\left\vert 1\right\rangle \}$ and $\{\left\vert
+\right\rangle ,\left\vert -\right\rangle \}$, the receiver Bob in their
protocol does not involve any nonorthogonal states nor noncommuting
operations, even though his operations are actually performed on quantum
systems. Such a usage of the term \textquotedblleft classical
Bob\textquotedblright\ is also adopted later in Ref. \cite{qi957}.

Similarly, in our protocol Bob's permutations $P_{i}$ are a group of
commutable operations which also exist in classical world. When using the practical implementation scheme in the
previous section, we can see that $P_{i}$ can be realized simply by
arranging the connection of the optical fibers between the input ports,
output ports and optical delays. Choosing different $P_{i}$ is basically the
same as resetting the jumpers in a classical telephone switchboard, and
Bob's role is very similar to a classical telephone operator. In a more
practical setting, instead of plugging and unplugging the optical fibers
manually, rearranging the connection between the input ports, output ports
and optical delays can be implemented efficiently using
microelectromechanical systems (MEMS) \cite{book11[10],qi2016} or other
optical cross-connect (OXC) devices \cite{qi2017,book11[8]}. Still, OXC
devices are also widely used in classical information optics, so that they
should be considered as classical despite that they can handle photons
carrying quantum information as well. Therefore, our protocol also takes a
\textquotedblleft classical Bob\textquotedblright\ only. Furthermore, our
Bob does not need the sources and detectors for preparing and measuring
quantum systems. In this sense, it is somehow more classical than the
\textquotedblleft classical Bob\textquotedblright\ in Refs. \cite%
{qi443,qi957}.

\section{Discussions}

In summary, we proposed an MDI-QKD protocol and considered its possible
implementation in practice. Comparing with previous proposals, we should
note that the original MDI-QKD protocol \cite{di22} has two distinctive
advantages: (i) all detector side channels are removed so that its security
does not need to rely on trusted measurement devices, and (ii) the secure
distance with conventional lasers can be twice as that of conventional QKD
while the key rate remains high in practice.

Ours does not have the advantage (ii) because in practical QKD nowadays, the
secure distance is limited by the distance between the optical sources and
the detectors. In our protocol (or conventional QKD such as the BB84
protocol), it means the distance between Alice and Charlie (or Bob), while
in the original MDI-QKD, both Alice and Bob send photons or coherent pulses
to Charlie so that the secure distance is the distance from Alice to Charlie
plus the distance from Charlie to Bob.

But the advantage (i) remains in our protocol, and it also has two more
advantages. First, it no longer needs joint measurements, and second, Bob
can be classical. It is worth studying whether there can be other practical
implementation schemes of our theoretical protocol so that Bob's filters and shutters can be further simplified, enabling more users with low technical capacity to enjoy the advantages of quantum cryptography in practice.

\section*{Acknowledgements}

The work was supported in part by Guangdong Basic and Applied Basic Research
Foundation under Grant No. 2019A1515011048.

\end{document}